\documentclass[preprint,showpacs,preprintnumbers,amsmath,amssymb,prb]{revtex4}

\usepackage{graphicx}
\usepackage{dcolumn}
\usepackage{bm}

\newcommand{\h}{$ B$}
\newcommand{\lo}{$\rm L_{0}$}
\newcommand{\q}{$Q_{\rm l}$}

\newcommand{\ef}{$B/\sqrt{P}$}

\begin{document}

  \title{Design of Q-Band loop-gap resonators at frequencies 34-36 GHz  for single electron spin spectroscopy in semiconductor nanostructures}

 \author{B. Simovi\v c}
 \affiliation{Laboratorium f\"{u}r Festk\"{o}rperphysik, ETH Z\"{u}rich, CH-8093 Z\"{u}rich, Schweiz}
\author{R.Schuhmann}
 \affiliation{Fachgebiet Theoretische Elektrotechnik, Universit\"{a}t Paderborn, 33098 Paderborn, Germany}
 
 \author{J. Forrer}
 \affiliation{Laboratorium f\"{u}r Physikalische Chemie , ETH Z\"{u}rich, CH-8093 Z\"{u}rich, Schweiz}

 \author{P. Studerus}
  \affiliation{Laboratorium f\"{u}r Festk\"{o}rperphysik, ETH Z\"{u}rich, CH-8093 Z\"{u}rich, Schweiz}
\author{S.Gustavsson}
  \affiliation{Laboratorium f\"{u}r Festk\"{o}rperphysik, ETH Z\"{u}rich, CH-8093 Z\"{u}rich, Schweiz}

\author{R. Leturcq}
  \affiliation{Laboratorium f\"{u}r Festk\"{o}rperphysik, ETH Z\"{u}rich, CH-8093 Z\"{u}rich, Schweiz}
\author{K.Ensslin}
  \affiliation{Laboratorium f\"{u}r Festk\"{o}rperphysik, ETH Z\"{u}rich, CH-8093 Z\"{u}rich, Schweiz}

\author{A. Schweiger}
 \affiliation{Laboratorium f\"{u}r Physikalische Chemie , ETH Z\"{u}rich, CH-8093 Z\"{u}rich, Schweiz}

 \date{\today}

 \begin{abstract}

We report on the design of loop-gap resonators
(LGR) operating in the frequency range 34-36 GHz with the goal to achieve single electron spin resonance (ESR) in quantum dot nanostructures.  We present a
comprehensive study of the magnetic field strength and the spatial
distribution of the electric and magnetic fields in the resonator
by means of  experiments and numerical simulations. 

\end{abstract}

 \pacs{73.23.Hk ,  73.63.Kv , 84.40.Dc}

\maketitle

\section{Introduction}

Manipulation and read-out of the quantum state of individual
electron spins in quantum dots (QD) prepared  in semiconductor
nanostructures, have been hailed as a promising avenue towards
quantum computation.\cite{Loss1998} Experimentally, single electron manipulation
and detection have become routine in nanostructures but the
feasibility of controlling individual electron spins is yet to be
demonstrated. Different approaches have been envisioned so far. The electrical manipulation and detection of spins via spin-orbit interaction has been considered theoretically.\cite{Levitov2003} It has been shown by optical means that electrical
modulation of the g-factor\cite{Salis2001,Kato2004} can induce
a rotation of the electron spins in GaAs parabolic quantum wells, but this method has not yet been successfully applied to QD.
\newline A more conventional approach is the technique of electron
spin resonance (ESR) which involves magnetic dipole transitions
induced by an oscillating magnetic field.  Observing ESR in QD  is
not without challenge, however,  for reasons that will be clarified
in the present work. The first difficulty lies in the smallness of the magnetic dipole of an individual electron spin. In "traditional" ESR
experiments, which are  dealing
with an ensemble of spins the microwave (mw) magnetic field $B$ is generated by a
resonating cavity. The magnetization is
measured via the electrical induction created by the time-dependent
magnetic dipole moments inside the cavity.  The minimum number of spins which can be detected this way is  about $10^8$ spins depending on the details of the apparatus and the g-factor.\cite{Poole1983} The ultimate limit of single spin detection was first reached for
individual paramagnetic centers in silicon, with a scanning
tunneling microscope\cite{Manassen1989} and a magnetic resonance force microscope\cite{Rugar2004} by taking advantage of the much more sensitive local interaction
between the spin and the scanning tip. Very recently, single spin detection has also been achieved for a single electron paramagnetic center in silicon from the time-resolved charging events of the electron trap. \cite{Xiao2004} Single spin manipulation has yet to be demonstrated.
\newline Unlike isolated paramagnetic defects in silicon, QD are purpose-designed
and fully tunable systems for which the number of electrons, the total
spin state and the coupling to external leads can be controlled by
modifying their electrostatic environment. This makes QD
versatile quantum devices but they have also an important drawback:
their electronic properties are very sensitive to electromagnetic
and electrostatic disturbances. The electronic properties of QD are also strongly temperature ($T$) dependent.   \newline A non-invasive method for detecting
individual spins in QD has been suggested by Engel and Loss who
have shown by calculations that the mw-driven Rabi oscillation of
an individual unpaired electron spin in a  QD causes additional 
current peaks in the sequential tunneling through the QD.\cite{Engel2001,Engel2002} The key idea is that the electrochemical potential of the QD containing an odd number
of electrons depends on the state of the unpaired spin under
the static magnetic field. A lower bound for the single spin-decoherence time,
the knowledge of which being so critical for quantum computation
schemes, could be measured from the spectral width of the
ESR-induced dc current.\cite{Engel2001,Engel2002}
\newline To understand the challenge in this experiment, two facts
about QD are important. First, the Zeeman energy splitting of the
unpaired electron spin must be larger than the thermal
broadening ($ 3.5 k_{\rm B} T$) of the current peaks observed
in sequential tunneling. Given the typical electronic temperature
of about 100 mK reached for QD in a standard $\rm ^{4}He-^{3}He$ dilution refrigerator and the g-factor of $\sim 0.4$ for electrons in GaAs, a
Zeeman energy of 125 $\mu \rm eV$ (i.e. 1.7 K) corresponding to a
Larmor frequency of 35 GHz seems reasonable. An important consequence of this
is that mw-induced heating must be kept low, a requirement not
easily fulfilled in this frequency range for a large mw magnetic
field. Second, electrical coupling between the electromagnetic
field and the QD must be reduced to a negligible level. This
effect produces spurious mw fluctuations of the electrochemical
potential of the leads with respect to the electrochemical potential
of the QD.\cite{VanderWiel} As a result, photon-assisted tunneling (PAT) will occur
regardless of the spin when the mw electric field ($E$) is strong.
\newline To excite individual spins in QD within these constrains, our
approach is to use loop-gap resonators (LGR). The large amount of mw magnetic energy
that can be stored in an LGR has been long recognized at X-band (8-12
GHz).\cite{Froncisz1984,Froncisz1984B,Wood1984,Hyde1989,Mehdizadeh1983,Piasecki1993} An LGR  is an open-shield resonator whose shape, reminiscent of lump circuits, provides a  spatial separation between $E$ and $B$ on distances significantly shorter than the mw wavelength. The first realization of an LGR operating at Q-band frequencies (34-36 GHz) was reported by Froncisz, Oles and Hyde\cite{Froncisz1986}. Based on this pioneering work, we further explore  the characteristics and  the performances of Q-band LGRs with the goal of making them suitable for experimenting on QD. The factors affecting their performances are studied and discussed in section. II. Measurements of the mw-induced transport through a QD in an LGR are described in section. III. Detailed numerical calculations of the distribution of $E$ and $B$ are shown in section. IV.  

 \section{The loop-gap resonator}

\subsection {Design}

The most basic LGR structure, sketched in Fig.~1, consists of two
loops of different sizes connected through a gap. Multi-gap LGR  structures \cite{Piasecki1993} are not considered here. The
body of the LGR  is made of copper and its structure is formed by electro-spark erosion of the metallic disc housed inside the cylindric shield that has a cut-off frequency which is a few GHz
higher than the resonance frequency to prevent radiation. Several LGRs were built for this experiment. Their characteristics are detailed in Table I. The mw power propagating as $\rm TE_{10}$ mode of a WR28 waveguide couples into the LGR through an elliptic iris with dimensions of  $0.55 \times $3 mm. The fine tuning of the impedance between WR28 and LGR is realized by a tuner, which consists of a teflon screw with a metallic top, housed in a 3-mm long piece of a waveguide WR28. The tuner is positioned in front of the coupling iris of the LGR. The typical return loss achieved with this coupler varied between -20 and -30 dB depending on details in the dimensions of the LGR and its sample load.
\newline Samples can be inserted in two ways. For standard bulk ESR
experiments, our test sample is made of a capillary quartz tube filled with
 $\gamma$-irradiated Ca formate powder (ICAF). The capillary quartz tube (with a diameter of $0.4 \times 0.3 \ \rm mm$) is inserted into the small loop (\lo) of the LGR from the top
through the cylindrical opening of the shield. For the QD experiment, the
heterostructure was inserted through the side access, a small
rectangular  window that provides access to the
top of  \lo\ (see Fig. 1, for more detailed Fig. 3) 
As will be demonstrated, the sample has to be positioned close to the edge of \lo\ to
minimize harmful interaction of the QD with $E$.
\newline The behavior of the LGR can be thought at first approximation as an extension of a lumped-circuit to the mw frequency range.\cite{Mehdizadeh1983,Hyde1989,Froncisz1986} The maximum $B$ is stored in the loop  \lo\  while the maximum $E$ occurs across the gap thereby providing a spatial separation between the two fields. The resonance frequency is expected to be 
independent of the length $h$ of the resonator, it depends mainly on
the dimension of the smaller of the two loops  and the
gap. The large loop  $\rm L_{1}$ ensures that electromagnetic energy is transfered  from the waveguide into the resonator and controls the magnetic return flux. \newline However,  experiments and numerical calculations of the electromagnetic field
distribution have shown that this simple picture is inadequate as
the wavelength is  comparable in size to  the LGR dimensions.\cite{Piasecki1993}
The dependence of the resonant frequency on the dimensions of the
LGR shows indeed a clear deviation  from the lumped-circuit behavior.\cite{Froncisz1982}
The dimension of the cylindric shield and the
length of the resonator were found to affect the resonant frequency. \cite{Froncisz1982}  Extensive numerical calculations  and test bench measurements  show a significant overlap between B and E at 10GHz.~\cite{Pfenninger1988,Forrer1996}We can therefore anticipate that sample loading and 
structural details (such as mechanical surface quality and shield geometry)  will become critical for the performance of LGR at 34-36 GHz. \newline Since the mw power has to be kept low to prevent heating of the QD, we first  determined  the structural changes,
which are most effective in increasing \h\ for a fixed incident mw power
$P_{0}$ and fixed resonance frequency $f_{0}$. Assuming that B is uniform in the volume $V_{\rm r}$ of the resonator, the most general relation between \h\  and $P_{0}$ is expressed as

\begin{equation}\label{1}
   B \propto \sqrt{\frac{Q_{\rm l} P_{0}}{f_{0} V_{\rm r}}},
\end{equation}

\noindent where  $Q_{\rm l}$
is the loaded quality factor of the resonator. \cite{Poole1983}  If we further assume by analogy with lumped-circuits that  most of the magnetic energy in the LGR is contained inside \lo, the volume $V_{\rm r}$  is  equal to  $\pi \phi_{0}^{2} h/4$. Since for resonators one can usually neglect radiation
losses compared to conducting losses, $Q_{\rm l}$ should scale as the 
volume-to-surface ratio. This means for LGR that $Q_{\rm l} \propto \phi_{0}$, which suggests that  one could reduce the length $h$ to decrease $V_{\rm r}$ and thereby enhance \h\, without altering \q. To investigate the validity of these assumptions, we have fabricated LGRs for different sets of structural parameters shown in Table I.  The length $h$ was downsized from 3 to 0.2 mm.  Fine readjustments of the whole structure were needed to keep the resonant frequency in the range 34-36 GHz. The addition of a sample window was expected to noticeably lower \q. The window is indeed at a distance of 400 $\mu \rm m$ from the edge of \lo\ and could potentially disturb the fringing fields nearby. Table I shows, however, that although variations of  \q\  are observed among the different LGRs, they do not  correlate with intentional modifications of the resonator shape. For this reason,  we believe that these variations are due to accidental differences in the quality of the electro-spark erosion.  It is therefore fair to conclude from Table I  that \q\ lies in a range which is remarkably robust against downsizing the length  of the resonator and opening the resonator shield for sample access. 

\subsection{Amplitude of the B-field in LGR}

The performances of the different LGRs were investigated by means of pulse ESR experiments on ICAF powder contained in a capillary quartz tube inserted into \lo. The experiments were done with a home-made Q-band high-power pulse bridge implemented with a Bruker ELEXSYS 380 console. The LGRs were critically coupled to ensure maximal directivity. Table I shows the amplitude of the spin echoes measured at room temperature in different LGRs. For a constant receiver gain, the duration of the two mw pulses were set to 80 and 160 ns, respectively, and the time delay between the pulses was  300 ns. This corresponds to a flip  angle of $90^{o}$ and $180^{o}$ for the electron magnetization for \h $=0.1 \ \rm mT$ since g=2 for electron spins in ICAF. Here one should keep in mind that the mw $B$-field measured from ESR absorption is half the value of the mw B-field in the LGR, since only half of the linearly polarized $B$-field drives transitions.\newline
The incident  mw power was adjusted to give the maximum amplitude for the spin echo. Fig. 2(a) shows the incident mw power, which maximizes the echo amplitude as a function of the length $h$. Experiments were done with LGRs with and without rectangular sample window. No significant differences between the two cases were observed. Evidently, Fig. 2(a) shows that the mw power decreases as a function of $h$ in reasonable agreement with what we expect from equation 1 (dashed line).
\newline Because the ESR signal at resonance increases with the mw power
absorbed by the sample, the signal amplitude decreases as the length $h$ is reduced (see Table I). If scaled to $h$, however, we observe that the signal amplitude remains comparable among the different LGRs, thereby indicating that the filling factor is reasonably independent of  $h$ within experimental accuracy. This makes LGRs very useful resonant structures for experimenting on rare and small samples. Unlike cavities whose shapes are constrained by the resonance frequency, we show here that the size of a LGR  can be matched to a large variety of  sample sizes while preserving its performances. 
\newline The filling factor, which relates to the mw field distribution inside the sample is irrelevant for QD. The key quantity is the frequency of the Rabi-oscillation of the individual electron spin driven by the peaked mw $B$-field assuming that the QD is positioned where $B$ is maximum.  It is indeed the Rabi-frequency which controls the amplitude of the ESR-induced dc current.\cite{Engel2001,Engel2002}.  This frequency can be measured with good accuracy at low temperature from the nutation curve of the magnetization in the ICAF sample. Measurements were done at $T$ =  20 K by integrating the free induction decay (FID) as a function of the mw pulse duration ($\tau_{\rm P}$) as shown in Fig. 2(b) for $h$ of  3 and 0.2 mm.  The fast damping of the nutation observed for $h$=0.2 mm  bears witness of the decrease in field homogeneity as h is reduced. 
The period of these oscillations is also a  measure of $B$ in the LGR.  Measurements done for different incident pulse powers provide a power efficiency of $\rm 0.032 \ mT.mW^{-0.5}$ for $h$ = 3 mm and  $\rm 0.1 \  mT . mW^{-0.5}$ for $h$ = 0.2 mm. The latter corresponds to a Rabi-frequency 
 of $\rm 560\ kHz.mW^{-0.5}$ for electrons in GaAs
(g-factor$\sim$ 0.4).\newline We also found revealing to compare the value of \h\ measured in LGR with the amplitude of the mw B-field in the rectangular waveguide ($ B^{\rm wg}$). The latter can be easily calculated and we found: $ B^{\rm wg} = 7.14 \times \ 10^{-4} \rm mT. mW^{-0.5}  $. The ratio $\zeta =  B/B^{\rm wg}$ is shown as a function of $h$ in Fig. 2(c). The rapid increase of  the ratio $\zeta$  as $h$ decreases  below 1 mm  shows  the large density of magnetic energy that can be achieved in LGR. They have therefore  an obvious advantage over open waveguides and antennas in providing locally a large mw B-field.

\section{ LGR and QD}

As we experiment with QD, both the sample inserted into the LGR and the detection scheme of the ESR absorption  are qualitatively different from what has been described previously. Problems  specific to this   experiment will occur.  In this section we first describe the sample and the loading procedure. We then show measurements of the mw-induced dc current  through the QD loaded into the LGR.  Our study  reveals how slight modifications of the sample environment  produce considerable differences in the effect of mw on the current through the QD. 

\subsection{Sample}

The QD was fabricated on a AlGaAs-GaAs heterostructure
which contains a two-dimensional electron gas (2DEG) 34 nm below
the surface (electron density $4.5 \times 10^{-15} \rm \ m^{-2}$
and mobility $25 \rm \ m^{2}(V.s)^{-1}$). A back gate situated 1.4
$\rm \mu m$ below the two-dimensional electron gas allows to tune
the electron density. An atomic force microscope (AFM) was used to
locally oxidize the surface of the semiconductor.~\cite{Fuhrer2002} As
shown in Fig. 3(a), the 2DEG is depleted below the oxide lines
defining a two-terminal QD (dotted circles in Fig. 3(a)) and
a nearby electrostatically coupled quantum-point contact acting as
a charge detector (not used here). Charge transport through the
QD is controlled via two sets of gates. Dc voltages are applied
to the gates G1, G2 to tune the tunnel barriers between the QD
and the leads S$_{\rm QD}$ and D$_{\rm QD}$. The plunger gate
combines P, S$_{\rm QPC}$ and D$_{\rm QPC}$ and sets the electron
chemical potential inside the QD. All measurements were carried
out in a $\rm ^{4}He-^{3}He$ dilution refrigerator with a based
temperature of 60 mK with the mw set-up.
 \newline Fig. 3(b) shows how the sample is loaded into the LGR. In order to couple the QD to the mw magnetic field while minimizing the load, the AlGaAs-GaAs wafer was cut as close as possible to the optically etched mesa structure on which the QD is patterned (see inset of Fig. 3b). The positioning of the
QD on the top of the small loop \lo\ is checked
with an optical microscope. Less than a 100-$\mu m$-slab of wafer
covers the loop. The minimum distance between the QD and
the loop is set by the electrical contact between the surface of
the LGR  and a 25-$\mu m$-thick Au wire stuck on the active
surface of the sample. \newline The mw are guided  through a 1.5-m-long WR28
rectangular waveguide which is made of 4 distinct sections.  The guiding
structure has been optimized to minimize the effect of impedance mismatch while keeping a reasonable electronic temperature of $~\rm 250 \ mK$.

\subsection{Dc transport through QD under mw irradiation}

Fig. 4 and 5 show the current through the QD  as a function
of the plunger gate voltage under continuous mw irradiation. The
measurements were done at zero magnetic field and the mw power at
source was varied from -40 to -5 dBm. Data shown in Fig. 4 were
obtained via  standard lock-in technique with an applied ac
bias voltage ($V_{rms} = 24 \ \mu\rm V$) between source and drain.
For the data of Fig. 5, dc detection without voltage bias was used
in order to measure the pumped current through the QD induced by mw.

Two distinct experiments are compared in Fig. 4 and 5. For both,
the GaAs-AlGaAs wafer was positioned as sketched in Fig. 3(b). In the first
experiment, the wafer and the ceramic part of the
sample holder were exposed to the electromagnetic field of the
LGR. The mw field is expected to be predominantly magnetic on the top
of \lo\  where the QD is located, but $E$
is significant in the extended area inside the metallic shield
of the resonator. Spurious coupling between the fringing E-field
in the extended area inside the metallic shield of the LGR and the 
QD may be enhanced  through the dielectric constant ($\epsilon = 12.9$) of GaAs 
and the large amount of dielectric material surrounding the QD. To investigate the
phenomenon, part of the wafer (edge) and the ceramic part of the
sample holder were coated with a 500-nm-thick Au layer (about
twice the skin-depth) in a second experiment to attenuate the
electromagnetic field inside the wafer. The Au layer and the back
gate were both set to ground potential. Only the surface of
the wafer facing the loop \lo\ was left
uncovered to allow coupling between the $B$-field and the QD.

 Fig. 4 shows sharp differences between the two experiments in the
power dependence of the current through the QD. In the case of
the unshielded wafer, a large broadening of the current peaks
develops for increasing incident power, while their base becomes more and
more asymmetric and distorted. The asymmetry becomes obvious in the 2D map of the current as a function of the incident mw power and plunger gate shown in inset of Fig. 4(a). These peculiar changes in the shape of the current peaks can be understood in terms of the PAT mechanism~\cite{VanderWiel} mentioned in the introduction.  The distortion is caused by satellite peaks in the dc current whose amplitude grows linearly against the main peak with the mw voltage induced across the QD. This effect combined with an increase of the electronic temperature (Joule heating) results in the disappearance of the Coulomb blockade as seen in Fig.~4(a). These observations contrast sharply with
the data for the shielded wafer. Indeed, broadening induced by mw is considerably less in that case and more importantly, no
significant asymmetry in the shape of the current peaks is
observed.\newline
To go further into the analysis, measurements of the mw-induced pumped current in the two
experiments are compared and juxtaposed with numerical simulations
in Fig. 5. The presence of a pumped current bears witness of some
asymmetry across the QD such as asymmetric capacitive coupling
with the mw $E$-field resulting in voltage bias mw fluctuations
($V_{\rm ac}$) or different temperatures between source and drain. The two mechanisms are sketched in the inset of Fig. 5.
Since the pumped current can be measured at zero bias voltage, it
is a very sensitive probe of the electromagnetic fields
surrounding the QD. In particular, the combination of asymmetric
heating and PAT can be simulated as described in ref~\cite{VanderWiel} and reveals distinctive features which would not appear if only one of the two contributions were effective. This is
illustrated in Fig. 5. For the unshielded wafer, the pumped
current undergoes qualitative changes with a continuous phase
rotation as power increases. Particularly striking is the
oscillatory behavior observed around -34 dBm. Yet these
particularities in the shape of the pumped current are clearly
absent for the shielded sample. As seen in Fig. 5, the distinctive
features of the data can be reproduced by simulation assuming the
combination of both asymmetric heating and PAT  for
the unshielded wafer while 'pure' asymmetric heating for the
shielded wafer is dominant.  The cause for asymmetry is not understood but likely to be connected to details in the shape of the QD circuit. Our data and analysis  strongly suggest  that electrical coupling is suppressed by the metallic shield on the wafer. Local perturbations of $B$ are also expected. Numerical simulations of the detailed field distribution surrounding the sample are needed to confirm the suppression of the electrical coupling and also determine the extent of the disturbance of the mw B-field.

\section{Calculation of the field distribution }

Our measurements of the mw induced dc current through the QD make clear that details in the field distribution surrounding the QD are important to consider. The field distribution inside the LGR can be calculated with a remarkably large precision by state-of-the-art techniques of finite-element resolution of the Maxwell equations. The simulations have been performed using the commercial tool CST Microwave Studio (MWS) [CST]. The method, the Finite Integration Technique (FIT), solves Maxwell  equations on a three-dimensional computational grid which is also used to represent the geometric details of the resonator. The simulations include the LGR as well as all coupling devices (iris, coupling screw and transition to waveguide) and the setups for sample access. Results are not only the electric and magnetic field distribution in the LGR but also the coupling behavior with a quantitative determination of external $Q$-values. One of the main challenges in these simulations is the high sensitivity of the resonance frequency of the LGR regarding small geometric details such as the gap width ($w$) and the gap length ($L$) (see Fig. 1). This requires a careful modeling of these regions with a sufficient number of mesh cells, which can be achieved either by an increased number of overall cells, or the application of so-called subgrid models. Parameter studies with different mesh resolutions have been performed to find a final mesh with acceptable numerical errors, defining a trade-off between accuracy and efficiency of the simulations. Moreover, numerical tests with varying (absorbing and closed) boundary conditions have shown that the radiation through the end planes of the cylindrical waveguide (shield) can be neglected (as expected). This is, however, not always true for the rectangular sample window, which must be carefully designed to prevent a considerable radiation of the energy stored in the resonator with an impact on both resonance frequency and $Q$-value. To include the Ohmic losses in the metallic parts of the structure, a special surface impedance model is used. It turns out that the losses in the gap region are mainly responsible for the loaded $Q_{\rm l}$ of the resonator at room temperature. The reflected waves at the WR28 waveguide are absorbed by a highly accurate waveguide boundary condition.\newline In a first time-domain simulation the structure is excited at the waveguide port by a normalized modulated Gaussian pulse with a bandwidth of 25 GHz to 40 GHz. The input and output signals at the port are monitored, and after a Discrete Fourier Transform (DFT) of these signals the reflection parameter of the structure can be obtained. Alternatively, the resonance frequency and quality factor can be extracted with high accuracy from the output signal alone using a Prony-Pisarenko-type spectral estimation technique, which has been re-implemented in Matlab for the postprocessing of MWS time-domain data. Finally, in a second time-domain run, the electric and magnetic fields at the obtained resonance frequencies are monitored using an online-DFT. The required CPU-time for both simulation runs (with between 50,000 and 150,000 mesh cells) is in the range of 10-30 min on a standard PC (2x 3.0 GHz Pentium CPU). Both sample access setups (the quartz tube and the rectangular window) have been modeled and simulated. All field plots are normalized to an incoming wave of 1 W peak power at the resonance frequency. The relevance of these calculations is supported by the agreement within 5\%  between the measured and calculated resonance frequency  for a given structure.

\subsection{LGR (h=3 mm) loaded with a capillary quartz tube}

We have seen that the length $h$ of the resonator can be tuned over a large interval connecting the two extremes: $ h>>\phi_{0}$ and $h<<\phi_{0}$. Although the precise limitations to these structural changes are yet to be determined, we expect as $h$ increases, that  the characteristics of the LGRs  approach the ones of gap-tuned resonant tubes originally invented by Blumlein.~\cite{Blumlein1941} The aim of this structures which have a large aspect ratio ($ h \sim \lambda$ and $\phi_{0} \sim \lambda / 30$) is to provide an intense and highly directional mw magnetic field, homogenous over the resonator length. This means a larger signal for standard ESR.  We also expect that the field distribution on the top of the loop \lo\  be less sensitive to the sample load. \newline The direction and homogeneity of the mw magnetic field have been determined by calculations  for $h$ = 3 mm (maximum length investigated) with a capillary quartz tube inserted into \lo. The results are shown in Fig. 6. The cross section is the ($x$, $z$) plane (see Fig. 1). From left to right, we see the coupling area which consists of the screw tuner housing,  the $\rm 2.8 \times 0.55 \ mm$ coupling iris  and the 2.2 mm loop $\rm L_{1}$. It is followed by the gap and the loop \lo\ filled with the capillary quartz tube. A large overlap between the inductive and capacitive parts of the LGR is evident in Fig. 6. The magnetic field propagates  through the gap while gradually decaying in amplitude.  The maximum mw electric field is observed across the gap but propagates into the sample area as confirmed experimentally by the downward shift of the resonance frequency  ($\delta f = -403 \rm MHz$ for $h$ = 3 mm)  induced by the capillary quartz tube ($\epsilon =4.7$).  It can be seen in Fig. 6 that the mw magnetic field is homogeneous over the full length of the loop \lo\ and its amplitude $B$ = $ \mu_{0} H (\rm A/m)$ is equal to $\rm 0.06 \ mT.mW^{-0.5}$. This value agrees remarkably well with the power efficiency measured in Section. II. B (note that we measure by ESR half of that field). A careful examination of the intensity contour lines (not shown) shows a drop of $B$ by  6\%   across the loop \lo, the maximum being found at the opposite edge from the gap. In a standard rectangular cavity operating in the $\rm TE_{102}$ mode at 35 GHz, the variation of amplitude of the mw magnetic field would be about 23\% along the same capillary tune ($z$ axis) tube based on the wave equation. An additional important feature is the gradient of the mw magnetic field at the top of \lo. We see in Fig. 6(b) that the mw magnetic field drops by 50\% within $\delta z =0.2 \  \rm Êmm$ above \lo. This defines an upper bound for the distance between the QD and the LGR along the z-axis. 

\subsection{LGR (h=3 mm)  loaded with QD sample}

The influence of the 500-nm-Au shield on the mw field surrounding the QD sample is investigated numerically. Fig.~7 shows the detailed distribution of the mw fields $E$ and $B$  inside the loaded LGR with $h$ = 3 mm. It is reasonable to expect that the influence of the QD sample on the mw field is mostly determined by the Au shield and the bulk dielectric material. The sample used in the calculations  is therefore reduced to a block of GaAs wafer of dimensions identical to the real heterostructure. Panels (a) and (b) present the field distribution for the unshielded sample. Panels (c) and (d) present  the field distribution for the shielded sample. The sample is positioned exactly as in the experiment (see Fig. 3).  The thickness of the shield on the wafer is, however,  $5 \mu \rm  m$ instead of 500 nm  due to the limited resolution of the computer model. The simulations show that the Au shield affects  much more the intensity of the fields than their orientation (see Fig.~8 for a vector representation). We therefore omit the orientation in Fig.~7 and focus on the gradient in amplitude inside the sample.  In agreement  with experiments,  we see in Fig.~7 that the mw field distribution near the QD is affected by the metallic Au shield. For the unshielded sample, the mw field $E$ which is oriented parallel to the 2DEG,  propagates into the sample. The amplitude of  $E$ at the QD is about 125 V.$\rm m^{-1} \ \rm mW^{-0.5}$. This is 1000 times less than across the gap but nevertheless comparable with the mw $E$ field in a standard WR28 waveguide operating at 35 GHz ($ E^{\rm WG} = 325 \ \rm V.m^{-1}$ for 1 mW). An electric field of this amplitude can easily disturb the dc voltage of the metallic gates controlling the QD. It can also cause intense Joule heating of the electrons (dielectric losses are negligible at low temperature). It is clear from Fig.~7(c) that these spurious effects must be suppressed for the shielded sample, for the $E$ field no longer penetrates into the sample. In Fig. 7(d), we see that the distribution of the mw magnetic field $B$ is also disturbed by the presence of the shield, becoming more inhomogeneous near the sample. Yet the amplitude and direction (not shown) of $B$ at the QD location are preserved which makes ESR possible with the sample shield. The variations of amplitude in the mw fields $E$ and $B$ between the two cases, unshielded and shielded sample, comes from different impedance matching conditions. The position of the coupler was indeed kept constant for all the calculations shown in this manuscript. Experimentally, however, the quality of the impedance matching changes with the sample load.  In particular, we  observed that the return loss increases to  about -10 dB (the value depends on the sample position) in the presence of the sample shield and the shape of the coupler had to be modified in order to reach  -20 dB.  This reduction of the transmitted mw power, caused by the Au shield, is in very good agreement with the reduction of the simulated mw fields in Fig.~7. 

\subsection{LGR ($h$ = 0.2 mm)  loaded with QD sample}

Given the increase in efficiency as the length $h$ is reduced, we have also studied numerically the mw field distribution in the LGR with $h$ = 0.2 mm  loaded with the shielded sample. The results are shown in Fig.~8.  As expected, both the amplitude and the direction of $B$ at the QD location are suitable for ESR experiment and the sample shield effectively prevents the $E$ field from propagating into the GaAs wafer. Unlike the 3-mm-long LGR however, we have observed experimentally that  the resonance frequency and the quality factor are strongly modified as we insert the shielded wafer.  Calculations predict  an upward shift  in frequency of 4 GHz with the sample load. Experimentally, we could no longer see the resonance coupling on the network analyzer. The shielded wafer, in contrast to the capillary quartz tube, is therefore too large a load for the 0.2-mm-long LGR.  Both the frequency shift and the losses caused by the load play a role in the reduction of the coupling  and mechanical adjustments are needed to counteract these effects. One way to restore the coupling is to lower the resonance frequency of the LGR by either filling the gap with dielectric material or modifying the gap geometry. We have measured the mw magnetic field $B$ with the gap half filled with teflon using a test sample consisting of  ICAF powder glued on a GaAs wafer. The experiment was done at low temperature ($T$ = 20 K) not to be influenced by dielectric losses. The teflon part was positioned at the maximum $E$-field i.e. next to $\rm L_{1}$ and the resonance frequency was 35.3 GHz in this condition. The wafer was positioned as depicted in Fig.~3. We achieved a return-loss of -5 dB which is far from optimal if compared to the -20 dB  measured with $h$ = 3 mm.  The LGR power efficiency for this load was measured by spin-echo as described  in section. II. B. We have found that despite suboptimal coupling, the efficiency \ef\  is a factor of 2 higher than for $h$ = 3 mm but a factor of 4 less than the one measured in the same LGR with a capillary quartz tube (see Fig.~2(a)). The reason for the loss of efficiency caused by the GaAs sample is not completely understood  but very likely connected with the size of the coupling iris which is no longer adequate for the loaded LGR. Work is in progress to optimize the shape and position of the coupling iris with the goal of improving the impedance matching between the waveguide and the resonator.

\section{Conclusion}

We have presented a detailed study of the performance of LGRs designed to operate at frequencies in the range of 34-36 GHz. The central motivation of the work is to observe ESR in QD. This experiment requires a large mw magnetic field in the QD while keeping the electronic temperature in the milliKelvin range and the mw electric field close to zero.  We have shown by means of experiments and numerical calculations that  LGRs can be designed to fulfill these constraints. The power efficiency reachs  $\rm 3.1 mT.W^{0.5}$ which is 20 times higher than a high-quality-factor $\rm TE_{102}$ cavity and 200 times higher than a waveguide. The spurious influence of the residual mw $E$ field on the QD such as Joule heating and capacitive coupling is reduced to a negligible level by a proper shielding of the dielectric materials surrounding the QD.  The remaining heating  caused by dissipation of the mw power into the resonator (heating power equals $ P_{0}/Q_{l}$ with $Q_{l}$=700-1000 at low temperature) is considerably reduced in a pulse experiment by limiting the pulse duty ratio. In that case, the ESR induced tunneling events of the electron through the QD can be accurately detected following a mw pulse by monitoring the jumps in the conductance of a QPC nearby as described in ref~\cite{Gustavsson}  (see also Fig. 3(a)). In principle, one should be able to measure the ESR spectrum from the QD  by counting the events detected on a QPC as a function of the static magnetic fields. This assumes that the correlation between the two events: applying a mw pulse and tunneling of an electron exists only at resonance.  Pulsed experiments were done at zero magnetic field with the shielded sample and a duty ratio sets to 1/1000. The pulse duration for a $\pi$-rotation of the electron spin is $\rm \tau_{\rm P}\approx 3 \mu s$ for 1 mW of pulsed power (LGR with $h$ = 3 mm). We observed that the background signal for the shielded sample consists of tunneling events uncorrelated with the mw pulse and whose statistics is entirely determined by the electronic temperature of 250 mK.  At this temperature the background signal is too large to permit ESR detection, the signal being blurred by the thermal broadening of the main conductance peak.  Work is in progress to lower the electronic base temperature down to 100 mK which would  be sufficient for ESR detection according to simulations based on ref.~\cite{Engel2001,Engel2002} and given the sensitivity of our detection scheme.~\cite{Gustavsson}

The authors thank M. Hausammann for his support during the fabrication process of the LGRs described in this work.

\newpage

\begin{figure}[t]
 \vspace{0cm}
\begin{center}
\includegraphics[width=0.5\linewidth, angle=0]{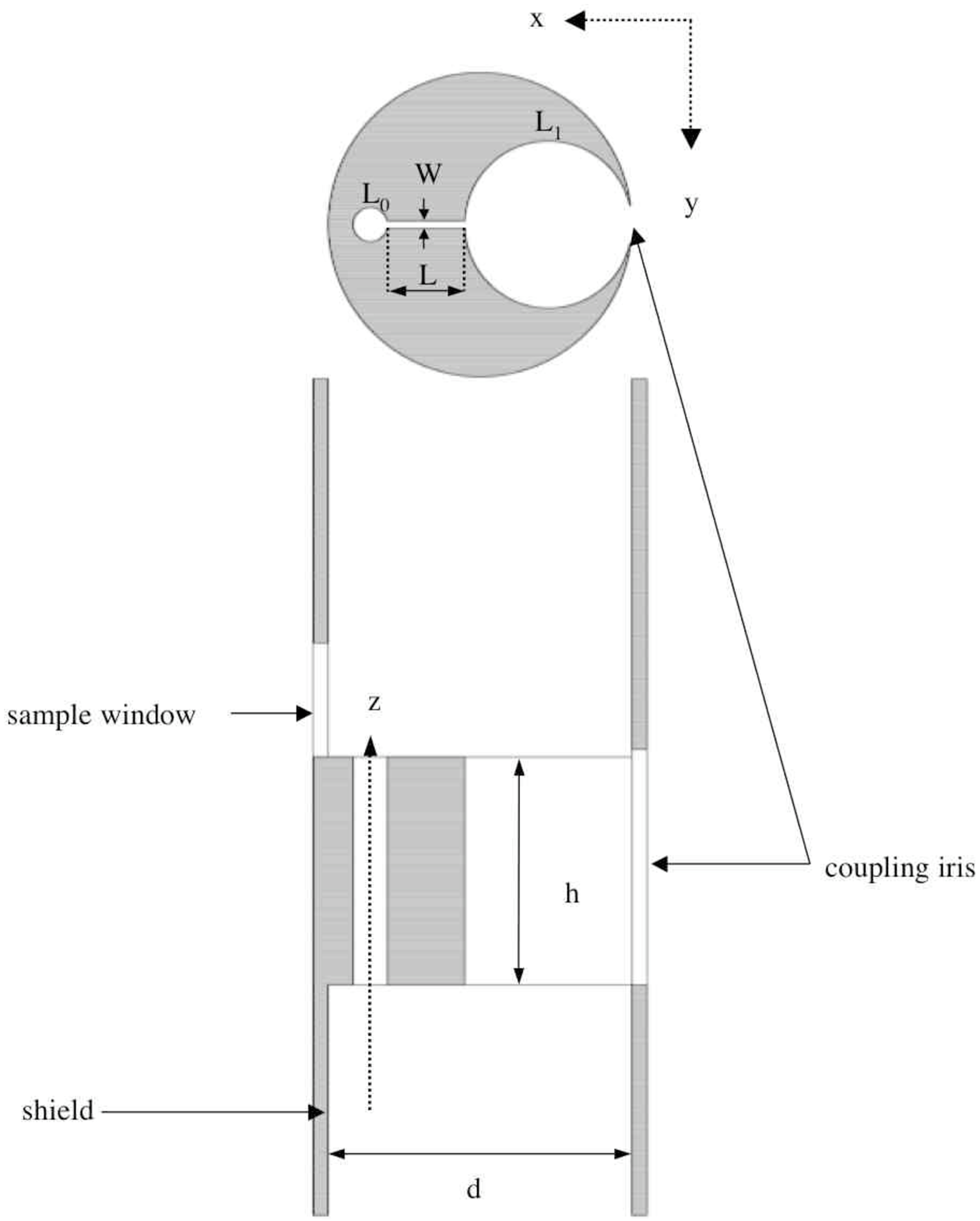}
\caption{Cross sectional views of the LGR  consisting of  two loops \lo\  and $\rm L_{1}$ connected through a gap $w$. The LGR structure is surrounded by a shield 
with diameter $d$ and a constant length of 18 mm. A small
window of dimension  $2 \times  1.5 \ \rm mm$ is made through the shield to
allow for a positioning of the QD on the top-edge of the small loop where
the magnetic field is expected to be maximum and the electric field minimum.
Microwave power coupling to the LGR is realized with a coupling iris. The
metallic structure as a whole is silver coated ($1\  \mu  \rm m$) to
reduce conduction loss. The LGR resonance frequency is gap tuned. For a  LGR of length  $h$ = 0.2 mm (Table I), a gap width ($w$) sensitivity of about 150 MHz/10 $\mu \rm m$ and a gap length ($L$) sensitivity of about 60 MHz/10 $\mu \rm m$ were found numerically. These values agree with experiments within 5\%.}
\end{center}
\end{figure}

\begin{table*}[t!]
  \centering
  \caption{Microwave characteristics of the LGR. The lines  marked with (w) show data for LGR with  a sample window (see Fig. 1). The different quantities $h$,   $L$, $W$, and $d$ are defined in Fig. 1.  The symbols $\phi_{0}$  and  $\phi_{1}$ stand for the diameter of \lo\ and $\rm L_{1}$, respectively. The signal is the peak value of a two-pulse (80/160 ns, $\tau$= 300 ns)  spin echo. We call $ Q_{l}$  the loaded quality factor measured in empty LGRs at room temperature  with a calibrated HP 8722 ES network analyzer.  }\label{h}
  \begin{tabular}{  c c c c c c c c c c c}
    \hspace{0.2 cm}\\
  \hline
  \hline
  \hspace{0.2 cm}\\
   $ $ & $h$ & $\phi_{0}$ & $\phi_{1}$ & $L$ & $W$ & $d$ & $f_{0}$ & $  $ & Signal & Signal/$h$ \\
   $ $ & (mm)& (mm) & (mm) & (mm)& (mm)& (mm) & GHz & $Q_{\rm l}$ & mV & mV/mm
  \hspace{0.2 cm}\\
  \hline
   \hspace{0.2 cm}\\

  $ $  & 0.35 & 0.48 & 2.0 & 1.0 & 0.126 & 4.5 & 35.6 & 174 & 6 & 17  \\
  $ $ & 0.6 & 0.48 & 2.0 & 1.03 & 0.16 & 4.5 & 35.35 & 195 & 10 & 16.7\\
  $ $ & 1 & 0.48 & 2.0 & 1.03 & 0.16 & 4.5 &35.0 & 210 & 18 &  18\\
  $ $ & 3 & 0.48 &  2.3 & 1.05 & 0.19 & 4.5 & 35.6 & 212 & 40 & 13
  \hspace{0.2 cm}\\
  \hline
  \hspace{0.2 cm}\\
  w & 0.2 & 0.45 & 2.3 & 1.15 & 0.10 & 4.1 & 34.95 & 316 &4& 20\\
  w & 3 & $0.39$ & 2.0 & 1.0 & 0.13& 3.9 & 38.0 & 220 & -- &--\\
  w & 3 & 0.48 & 2.0 & 1.0 & 0.13 & 3.9 & 36 & 180 & 25& 8 \\
  \hline

  \hline
\end{tabular}
\end{table*}

\begin{figure}[t!]
 \vspace{0cm}
\begin{center}
\includegraphics[width=0.4\linewidth, angle =0]{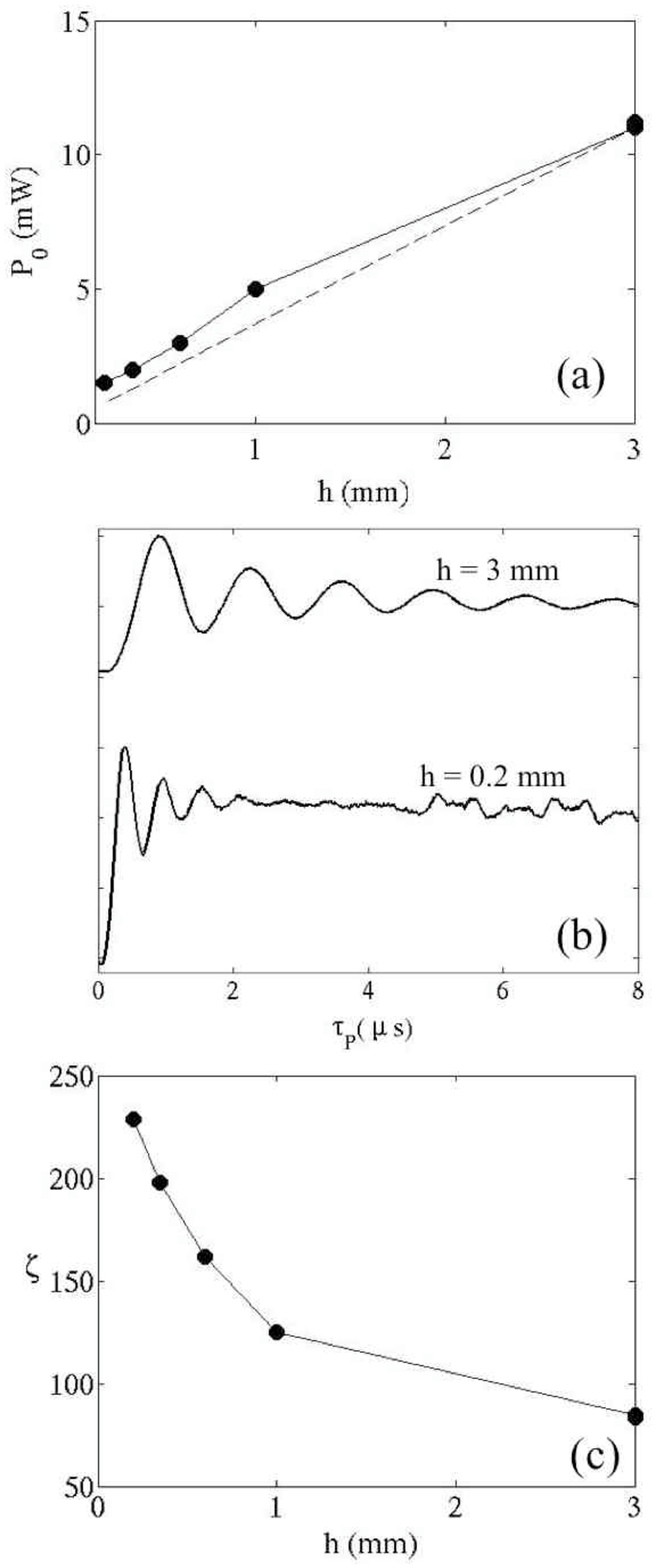}
\caption{(a) Incident pulsed mw power in LGRs as a function of the length
$h$ of the resonator for a field strength (rotating component)  \h = 0.1 mT in the sample. The dashed line corresponds to the calculated behavior. (b) Nutation curves of the magnetization measured at the temperature $T$ = 20 K
on a powder sample of ICAF contained in
a thin capillary quartz tube placed in the mw field inserted into  \lo. The pulsed mw power is -2.2 dBm (600 $\mu \rm W$) for $h$ = 3 mm and -3.9 dBm (400 $\mu \rm W$) for $h$ = 0.2 mm. This corresponds to  mw field strength of 0.025 mT and 0.062 mT, respectively.  The two LGRs have a window for sample access. (c) The ratio $\zeta =  B/B^{\rm wg}$ as a function of the length $h$ (see text for explanations).}
\end{center}
\end{figure}

\begin{figure}[t!]
 \vspace{0cm}
\begin{center}
\includegraphics[width=0.85\linewidth, angle=0]{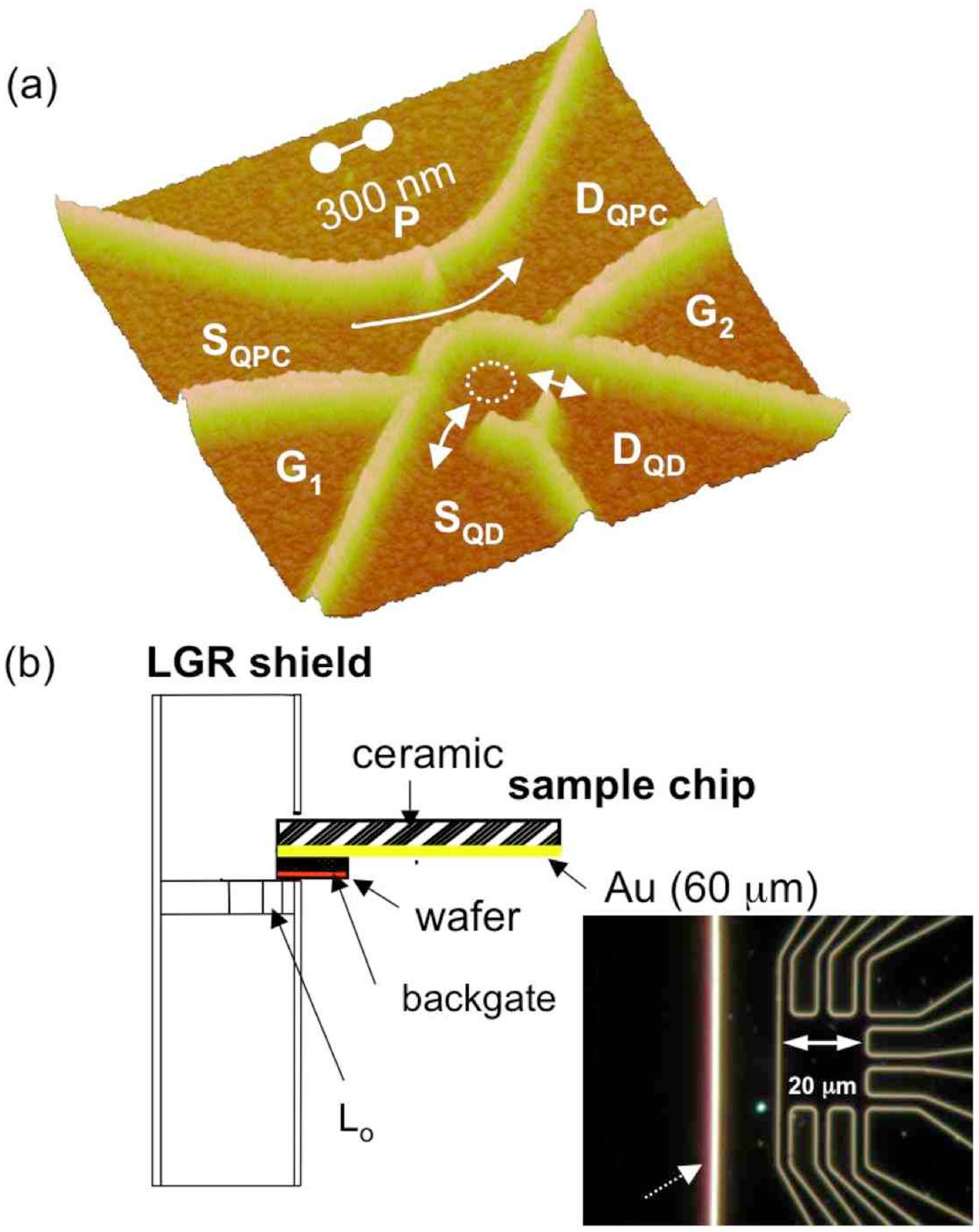}
\caption{(a)AFM-defined QD used in the experiment.
Charging energy and single-particule energy-level spacing, all inferred
from the measurement of the Coulomb diamonds (not shown) are:
$\sim 2 \ \rm meV$  and 200$\rm \ \mu eV$, respectively. (b)
Sketch of the GaAs-AlGaAs wafer mounted on a ceramic chip and inserted through the metallic shield of the LGR. The sample holder, which allows ($x$, $y$) and $z$ positioning is not shown on this figure.
Inset: picture of the wafer cut  (the edge is indicated by the dotted arrow) near the optically etched structure supporting the AFM defined QD.}
\end{center}
\end{figure}

\begin{figure}[t!]
\vspace{1cm}
\begin{center}
\includegraphics[width=0.5\linewidth, angle=0]{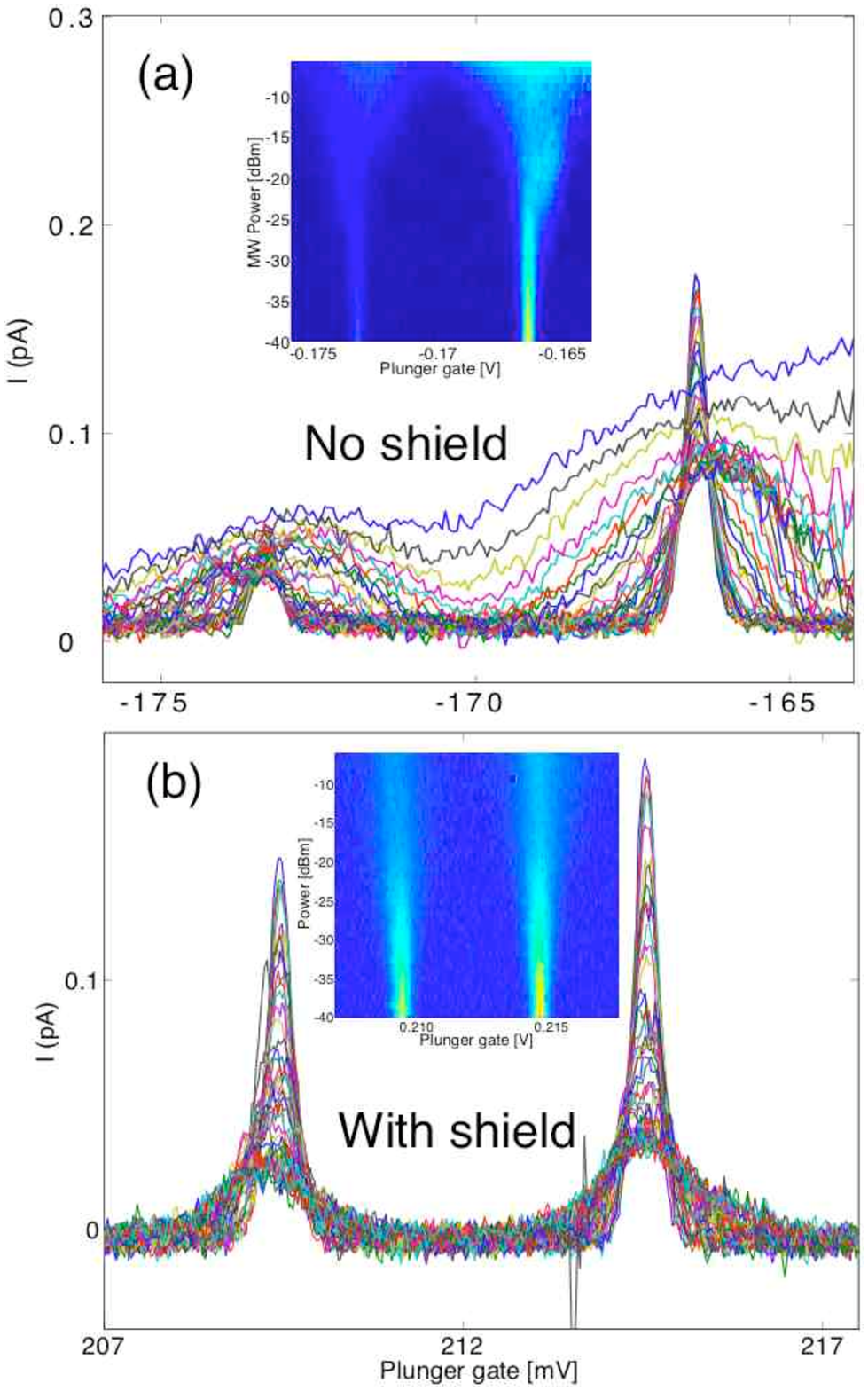} \caption{Current through the QD as a function of gate voltage and the incident mw power. The power at source was varied from -40 to -5 dBm. The attenuation through the waveguide is $\sim$ 2.5 dB.
(a) The  GaAs-AlGaAs wafer and the ceramic part of the sample holder are
not shielded against the electromagnetic field. (b) The edge sides of the GaAs-AlGaAs wafer
and the ceramic part of the sample holder are covered by a
500-nm-thick Au layer which is set to ground potential.}
\end{center}
\end{figure}

\begin{figure}[t!]
\includegraphics[width=1\linewidth, angle = 0]{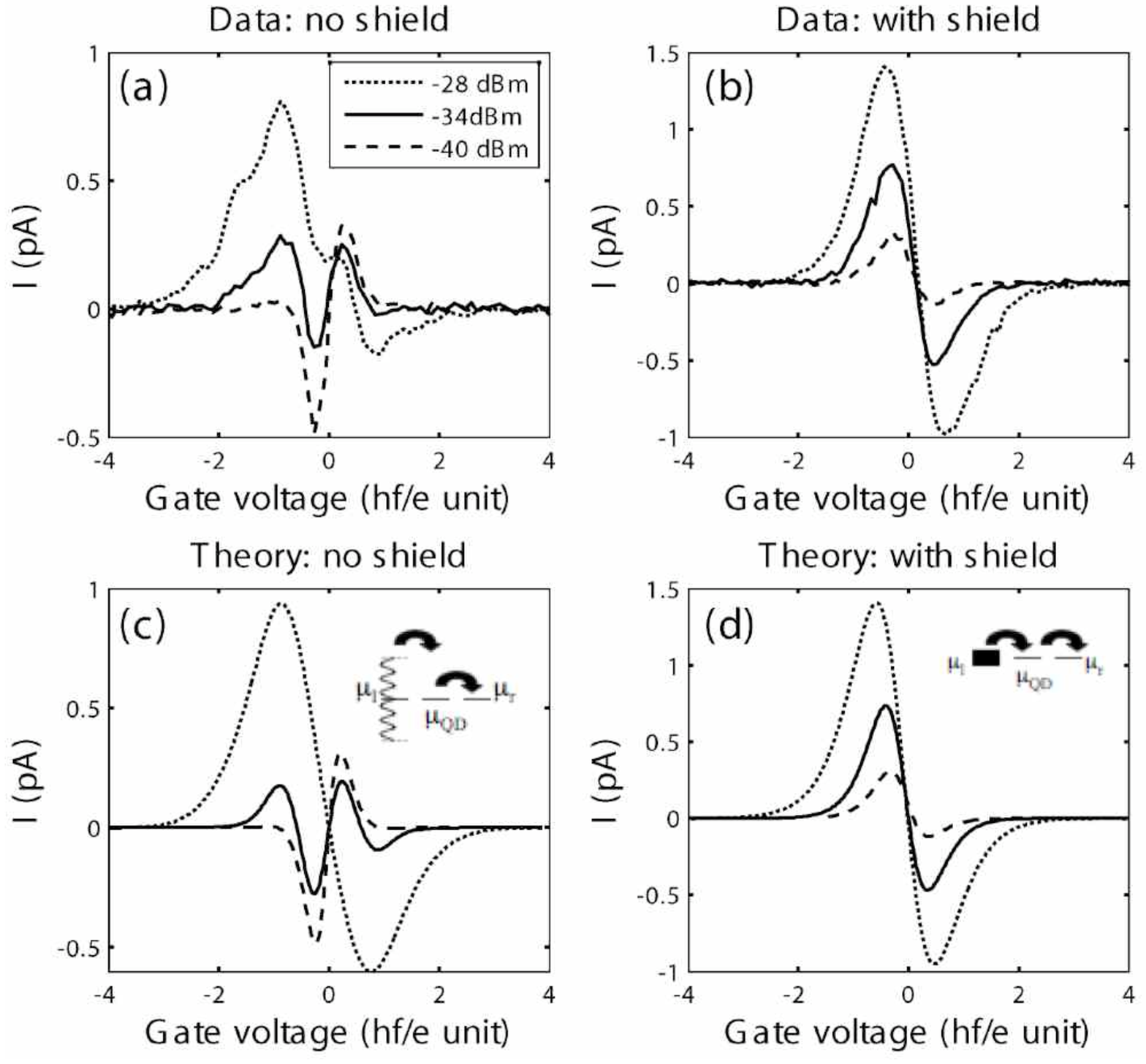} \caption{Influence of the Au layer covering the GaAs-AlGaAs wafer on the pumped current through the QD  under mw irradiation.
 Measurements (panels (a) and (b)) and simulations (panels (c) and (d)) are  compared for three different
 incident mw powers. The parameters of the simulations for -28, -34 and -40 dBm  are for
(c) unshielded wafer: $T_{l}^{\rm lead}$ = 788, 388, 270 mK,
$T_{r}^{\rm lead}$ = 640, 330, 235 mK  and $V^{\rm ac}_{\rm r}$ =
155, 52, 15 $\mu \rm eV$ and for (d) shielded wafer: $T_{\rm l}^{\rm
lead}$ = 546, 407, 370 mK , $T_{\rm r}^{\rm lead}$ = 628, 437, 379
mK  and $V^{\rm ac}_{l/r} = 0$. Insets: sketched of the mechanism for pumped current through the QD. On the left inset, the chemical potential of the left lead ($\mu_{\rm l}$) oscillates up and down at the mw frequency in respect to the chemical potentials of the QD ($\mu_{\rm QD}$) and the right lead ($\mu_{\rm r}$), respectively. As we continuously sweep $\mu_{\rm QD}$, the dc current changes sign when all three chemical potentials are equals. On the right inset, $\mu_{\rm l}$  no longer oscillates but widens due to the temperature increase in the lead. \label{fig:samplein}}
\end{figure}

\begin{figure*}[t!]
 \vspace{0cm}
\begin{center}
\includegraphics [width=1\linewidth, angle=0]{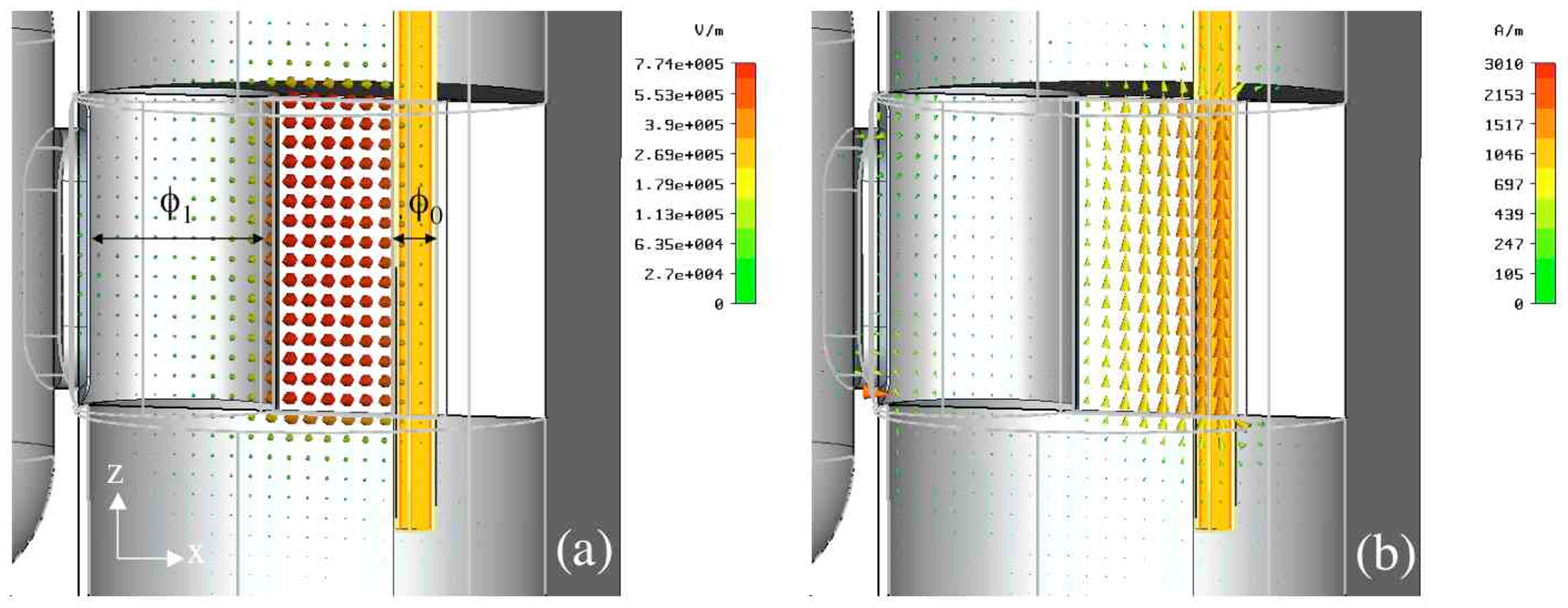}
\caption{Cross sectional ($x$, $z$) views of the spatial distribution of $E$ (panel (a)) and $B$ (panel (b)) in a LGR for $h$ = 3 mm. The color scale is logarithmic. The LGR  is loaded with a capillary quartz tube (orange) inserted into \lo\ through the cylindrical shield.  The LGR structure is without sample window (see Fig. 1). $\phi_{0} = 0.45 \mu \rm m$ and $\phi_{1} = 2.1 \rm mm$ stand for the diameter of the loops \lo\ and $\rm L_{1}$, respectively.}
\end{center}
\end{figure*}

\begin{figure*}[t!]
 \vspace{0cm}
\begin{center}
\includegraphics [width=1\linewidth, angle=0]{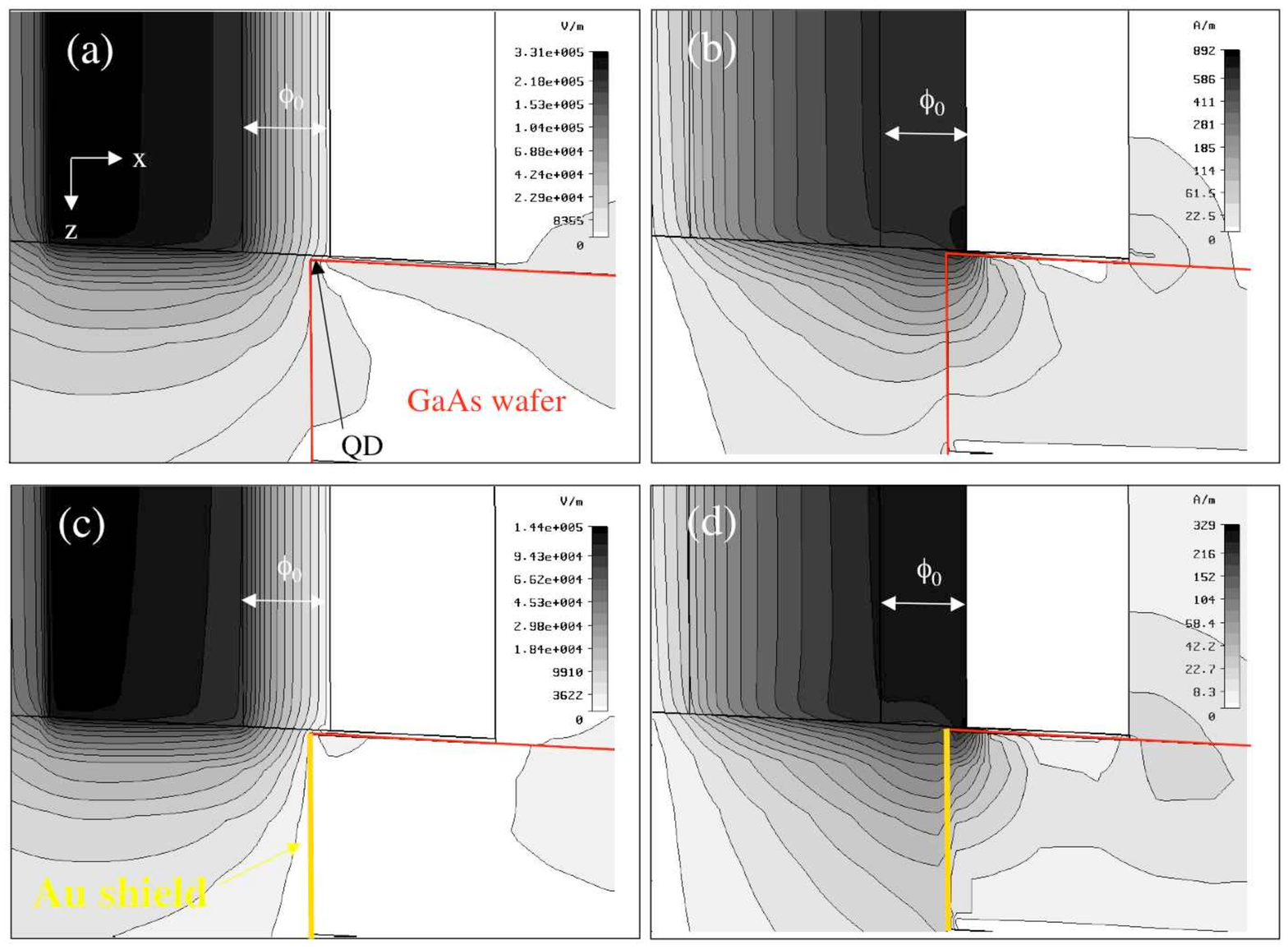}
\caption{Cross sectional views ($x$, $z$)  of the distribution of $E$ and $B$-field amplitude in the gap and the loop $\rm L_{0}$Ê for a LGR with length  $h$ = 3 mm and loaded with a GaAs wafer. The boundaries of the wafer are underlined in red. The color scale of the contour plot is logarithmic. Panels (a) and (b) show the distribution of $E$ and $B$, respectively for the  unshielded wafer. Panels (c) and (d) show the distribution of $E$ and $B$, respectively for the shielded wafer. The shield in the simulation is a 5 $\rm \mu m$-thick Au layer indicated by the yellow line. $\phi_{0} = 0.45 \mu \rm m$ is the diameter of \lo.}
\end{center}
\end{figure*}

\begin{figure*}[t!]
 \vspace{0cm}
\begin{center}
\includegraphics [width=1\linewidth, angle=0]{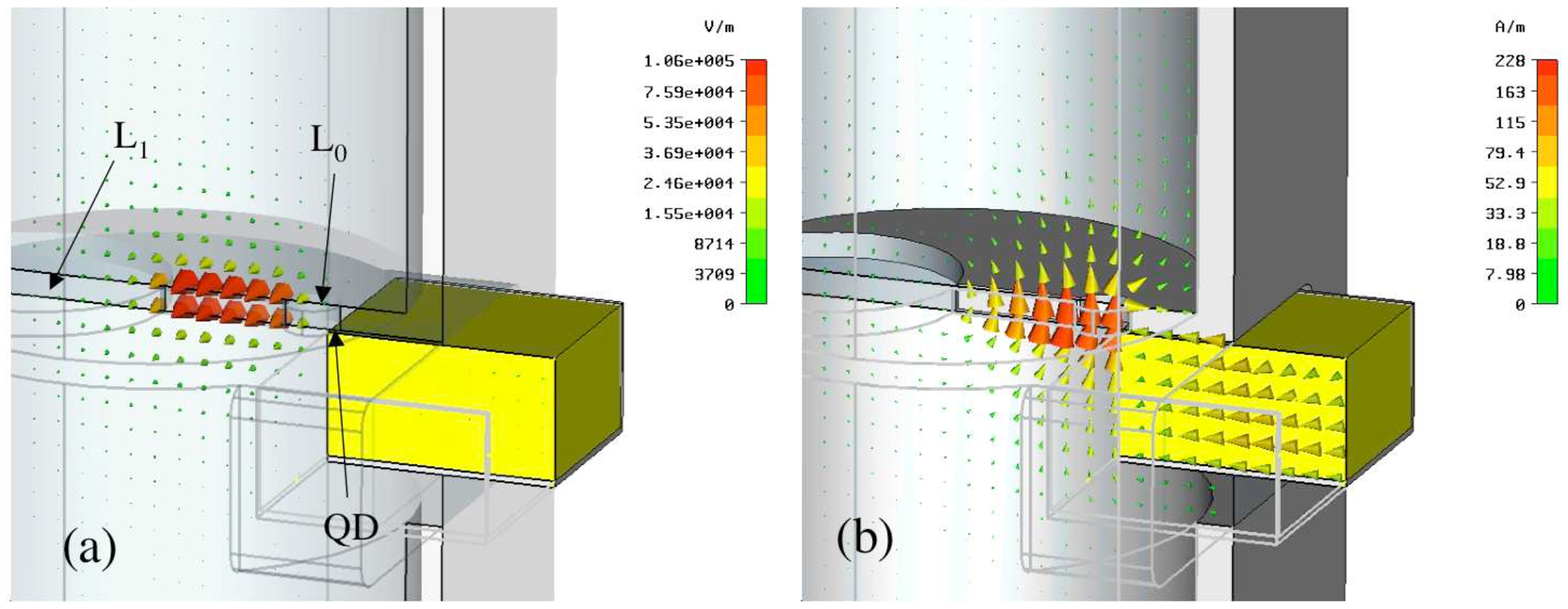}
\caption{Cross section of the 0.2-mm-long LGR loaded with the shielded GaAs wafer (yellow).  Note that the Au shield is not resolved on the scale of this figure. Panels (a) and (b) show the spatial distribution of $E$ and $B$, respectively. The color scale is logarithmic.}
\end{center}
\end{figure*}

\end{document}